# Passive phloem loading and long-distance transport in a synthetic tree-on-a-chip.

Jean Comtet[1,*], Kaare H. Jensen[2], Robert Turgeon[3], Abraham D. Stroock[4,5], A.E. Hosoi[1]

1. *MIT Mechanical Engineering, Cambridge, Massachusetts 02139, USA*
2. *Department of Physics, Technical University of Denmark, DK-2800 Kgs. Lyngby, Denmark*
3. *Section of Plant Biology, Cornell University, Ithaca, NY 14853, USA*
4. *School of Chemical and Biomolecular Engineering, Cornell University, Ithaca, NY 14853, USA*
5. *Kavli Institute at Cornell for Nanoscale Science, Cornell University, Ithaca, NY 14853, USA*
\* *Correspondence: jean.comtet@gmail.com*

## Abstract

Vascular plants rely on differences of osmotic pressure to export sugars from regions of synthesis (mature leaves) to sugar sinks (roots, fruits). In this process, known as Münch pressure flow, the loading of sugars from photosynthetic cells to the export conduit (the phloem) is crucial, as it sets the pressure head necessary to power long-distance transport. Whereas most herbaceous plants use active mechanisms to increase phloem concentration above that of the photosynthetic cells, in most tree species, for which transport distances are largest, loading seems to occur via passive symplastic diffusion from the mesophyll to the phloem. Here, we use a synthetic microfluidic model of a passive loader to explore the nonlinear dynamics that arise during export and determine the ability of passive loading to drive long-distance transport. We first demonstrate that in our device, phloem concentration is set by the balance between the resistances to diffusive loading from the source and convective export through the phloem. Convection-limited export corresponds to classical models of Münch transport, where phloem concentration is close to that of the source; in contrast, diffusion-limited export leads to small phloem concentrations and weak scaling of flow rates with the hydraulic resistance. We then show that the effective regime of convection-limited export is predominant in plants with large transport resistances and low xylem pressures. Moreover, hydrostatic pressures developed in our synthetic passive loader can reach botanically relevant values as high as 10 bars. We conclude that passive loading is sufficient to drive long-distance transport in large plants, and that trees are well suited to take full advantage of passive phloem loading strategies.

# Introduction

Sugars, photosynthesized in the mesophyll of plant leaves, are conveyed by the sap to regions of growth or storage (such as roots or fruits) through a specialized vascular network called the phloem. In this microfluidic network, sugars play a dual role, acting both as an energy carrier and a motile force generator, where the osmotic pressure of the phloem sap provides the necessary force to drive a convective flow from sugar sources to sinks (Fig. 1a). Despite consensus over the basic principles governing sugar transport in plants, it remains unclear whether expenditure of biochemical energy is necessary to actively raise sugar concentration in the phloem above that of the photosynthetic cells (1) and whether transport of sugars in plants can be osmotically powered over long distances (2). Improved understanding of the physiology of phloem transport could also lead to the design and development of new classes of osmotically-driven microactuators (3).

The mechanism by which plants transport sugars from the mesophyll (Fig. 1a; green) to export conduits (Fig. 1a; red), known as phloem loading, is crucial. The loading of sugars lowers the chemical potential of water in the phloem cells, allowing water to enter the phloem via osmosis from the adjacent xylem (Fig. 1a, blue), and to be subsequently exported via the transport phloem. Phloem loading thus sets the pressure head available to power long-distance transport. In most plants, energy is used to increase the sugar concentration in leaf phloem above that of the mesophyll cells, which has traditionally been thought a requirement to overcome viscous drag in the sieve tubes and drive long-distance transport via Münch pressure flow (4). Because transport resistances scale with the height of the plant, one might expect trees to require the largest phloem pressures of all plants in order to sustain comparable rates of sugar export. Oddly, the majority of trees seem to rely on a passive phloem loading mechanism, in which the sugar concentration in the phloem is slightly lower than that of the photosynthetic cells (Fig. 1a, concentration diagram), thus reducing the apparent available driving force (5). Some experiments have provided evidence for the use of passive loading in trees (6, 7). Fu et al. (8) suggested that the low water potential in leaf xylem of trees requires large sugar concentrations in the mesophyll cells, making active phloem loading unnecessary, but this factor is not directly coupled to the kinetics of phloem transport. Hence, the ability of the passive loading mechanism to generate sufficient pressure to drive long-distance transport in plants is uncertain and the reasons for absence of active loading in trees are not well understood.

To explore whether passive phloem loading can generate the pressures necessary to drive sugar translocation in large trees, and to determine why trees might favor this mechanism, we investigate transport dynamics in a synthetic microfluidic system in which sugars are passively loaded - a so-called tree-on-a-chip (Fig. 1). Experimental systems designed to mimic transport processes in plants have been devised by several authors to test mechanistic hypotheses of vascular physiology (9–19). Existing synthetic designs of phloem transport, however, do not reproduce the interactions thought to exist between the phloem tube, the photosynthetic cells and the xylem, and are limited by the fact that the solutes present in the source are depleted by convection over time, preventing the development of steady flows and pressures over experimentally relevant timescales (9, 10, 12, 13, 16, 17, 19). Our synthetic tree-on-a-chip (Figs. 1b-d) allows us to overcome these challenges and replicate the dynamics of the passive loading mechanisms under conditions that range from small herbs to large trees. Due to the presence of a large sugar source reservoir, this bioinspired osmotic pump can run at steady-state for several hours or more, allowing precise measurements of the physicochemical coupling at the system scale. It may also provide new strategies for pumping microscale flows in lab-on-a-chip applications (20–25).

Using our synthetic passive loader, we study in this paper the full dynamics of passive phloem loading, set by the nonlinear interplay between advection out of the phloem and diffusion from the source, and demonstrate the existence of two limiting export regimes. We analyze the efficiency of these regimes with respect to sugar and water export and their dependence on both local and global physiological plant traits. We then consider whether it is feasible for passive loading mechanisms to generate the hydrostatic pressures necessary to drive and sustain long-distance transport in plants and tie our results back to the predominant use of passive phloem loading among tall trees.

**Results**

*Experiments on sap flow in a tree-on-a-chip*

Sugars synthetized in the mesophyll cells are exported by bulk osmotic flow of water through the phloem conduits (Figs. 1a and b). We consider in Fig. 1b a simplified model of a passive symplastic loader. Sugars (red dots) first diffuse (red horizontal arrows) from the mesophyll source cells (S, green) to the phloem (P, pink) through passive intercellular channels called plasmodesmata (porous interface). The presence of sugars in the phloem drives an osmotic flow of water (blue arrows) from the adjacent xylem tissue (X, blue). This osmotic flow increases pressure in the phloem, and drives a bulk flow of water and sugars out of the leaf through the transport phloem (red and blue downward arrows).

We designed and fabricated an osmotic micropump that captures the main physical ingredients of sugar translocation in a passive phloem loader (Fig. 1c). The phloem channel (P, pink), connected to the pump outlet is in contact with a large sugar source (S, green) via a porous interface (red) that allows sugars to diffuse in. The phloem channel is also in contact with the xylem water source (X, light blue) via a semipermeable membrane (blue) that allows for osmotic exchange of water between the two channels. The photo in Fig. 1d shows the actual device.

During a typical experiment, we first flush all channels with pure water, then introduce a dextran[1] solution of concentration $c_0$ to the mesophyll source reservoir. To measure the flow rate $Q(t)$ exiting the synthetic phloem, the inlet and outlet of the device are connected to partially filled glass capillary tubes. Initially a transient phase is observed in which the flow rate increases with time due to a gradual buildup of sugars in the phloem. Following this transient, stable osmotic pumping is observed (Fig. 2, inset).

The steady state flow rate is determined by a balance between diffusion and advection. Diffusive transport of sugars from the mesophyll to the phloem tends to increase the concentration in the phloem. By contrast, osmotically driven flow of water from the xylem out of the phloem flushes away sugars and reduces concentration in the phloem. Our device thus exhibits a qualitatively different behavior than that observed in previous experiments (12, 13, 16–18, 26, 27), where solutes present in the source were depleted over time, preventing the creation of steady flow over experimentally relevant timescales. In contrast, our device captures essential aspects of the dynamic interaction that is thought to exist between the phloem tube, the photosynthetic cells and the xylem.

---

[1] Dextran plays here the role of sugar in a live plant

*Effects of source concentration, loading conductance, and transport resistances*

To identify the parameters that influence sugar transport in passive phloem loaders (Fig. 1b), we systematically varied the source concentration $c_0$ [mol/m$^3$], the diffusive loading conductivity $K_D$ [m$^3$/s] between the mesophyll and the phloem, the hydraulic semipermeable membrane resistance $R_M$ and the hydraulic transport resistance $R_T$ [Pa.s/m$^3$]. The loading conductivity is characterized by the coefficient $K_D$ [m$^3$/s] = $A \cdot k_D$ with $A$ [m$^2$] the membrane area and $k_D$ [m/s] the loading conductance per unit area. We denote the total hydraulic resistance of the system $R_{tot} = R_M + R_T$ (see *Materials and Methods*). All physical parameters and corresponding notations are summarized in Table 1.

Fig. 2 shows that the flow rate $Q$ [m$^3$/s] increases with the mesophyll source concentration $c_0$ and with the mesophyll-to-phloem loading conductance $K_D$ (red symbols correspond to similar total hydraulic resistance $R_{tot}$, but loading conductance $K_D$ is larger for red dots than for red triangles). Conversely, flow rate decreases with increasing hydraulic resistance (all dots correspond to similar loading conductance $K_D$ but hydraulic resistance is larger for green dots than red dots). The solid lines in Fig. 2 represents the flow rates that are expected when the phloem concentration is equal to the source concentration ($c = c_0$), as is assumed in classical models of Münch transport, for which $Q_M \sim RTc_0/R_{tot}$. We observed that the flow rates can deviate significantly from this prediction, since the finite loading conductance $K_D$ of the porous plasmodesmata-like interface reduces phloem concentration relative to that in the source ($c < c_0$).

*Theoretical analysis: Flushing number and transport regimes*

We now estimate how export in our synthetic passive loader deviates from the standard Münch model, for which phloem and source concentration are equals ($c = c_0$, solid lines in Fig. 2). Using Fig. 1b, we develop a mathematical model of passive phloem loading, following the model of symplastic phloem transport proposed by Comtet et al. (28). The sugar concentration $c$ [mmol/L] in the phloem chamber (of volume $V$ [m$^3$]) increases due to diffusion from the mesophyll and decreases by convective flow out of the phloem according to the conservation equation: $V\frac{dc}{dt} = K_D(c_0 - c) - Q \cdot c$. The first term on right-hand side is the diffusive transport of solutes across the porous barrier between mesophyll and phloem (driven by the concentration difference $c - c_0$), and the second term is the advective flow of solutes out of the phloem (equal to the mass flow rate of water through the phloem $Q$, times phloem sugar concentration $c$). When the system has reached steady state (Fig. 2, inset, horizontal dashed line), the balance of sugar concentration is expressed as:

$$\mathbf{K_D(c_0 - c) = Q \cdot c.} \qquad (1)$$

Eq. (1) characterises the balance between diffusive transfer between the mesophyll and phloem (left side term) and convective export of solute out of the phloem (right side term).

The phloem water flow rate $Q$ going from the xylem to the roots (blue arrows, Fig. 1b) is driven by the osmotic pressure in the phloem, and can be expressed as:

$$R_{tot}Q = RTc + P_X - P_R, \qquad (2)$$

where the right side of Eq. (2) represents the total driving force for the flow, with $RTc$ the osmotic pressure in the phloem, $P_X$ the xylem pressure, $P_R$ the root hydrostatic pressure (Fig. 2b) and $R_{tot} = R_M + R_T$ is the total hydraulic resistance of the system. Physical parameters and corresponding notations are summarized in Table 1. Deviation from Van't Hoff law can be taken into account as shown in *SI.1*.

To characterize the coupling between diffusive loading into the phloem chamber and advection out, we introduce a non-dimensional "flushing number" $f$ that represents the relative importance of these two processes:

$$f = \frac{ADVECTION}{DIFFUSION} = \frac{RTc_0 + P_X - P_R}{K_D R_{tot}} = \frac{Q_M}{K_D}. \qquad (3)$$

$Q_M = (RTc_0 + P_X - P_R)/R_{tot}$ [m³/s] is the flow rate obtained in the Münch scenario where phloem and source concentration are equal ($c = c_0$). $K_D$ [m³/s] is the loading conductance. Note that the flushing number is analogous to a system-scale (as opposed to a local) Peclet number, dependent on the physical parameters of the system. The representation of the system's dynamics by this single parameter is valid in our experimental conditions as well as in plants (see *SI.1*).

To illustrate the usefulness of the flushing number in characterizing the dynamics of the system, we can solve Eqns. (1-2) for the flow rate $Q$ and the concentration $c$:

$$\frac{Q}{Q_M} = \frac{c}{c_0} = \frac{\sqrt{1+4f}-1}{2f}. \qquad (4)$$

Dotted lines in Fig. 2 are solutions of Eq. (4) and match well our experimental datas.

We can also isolate the sugar export rate $\phi$ [mmol/s] (the product of flow rate and concentration) and rescale it by the maximal export rate achievable in the system $\phi_M = Q_M \cdot c_0$, obtained when phloem and source concentrations are equal:

$$\frac{\phi}{\phi_M} = \frac{Qc}{Q_M c_0} = \frac{(\sqrt{1+4f}-1)^2}{4f^2}. \qquad (5)$$

Note that this ratio, which we define as the Münch efficiency, depends only on the magnitude of the flushing number $f$.

*Experimental validation*

The flow rates measured in our synthetic passive loader are in qualitative agreement with our theory (dotted lines, Fig. 2). Moreover as shown in Fig. 3a, the flow rates collapse onto a single curve when rescaled by the maximal Münch flow rate $Q_M = RTc_0/R_{tot}$, as predicted by Eq. (4). The solid line in Fig. 3a is the theoretical expression from Eq. (4). This collapse gives us confidence that our model can capture the nonlinear transport dynamics arising in passive phloem loading over a broad range of resistances, loading conductivity and source concentrations.

We observe two regimes as a function of the flushing number $f$ in Fig. 3a, which are schematically represented in Figs. 3b-d. At low flushing numbers ($f \ll 1$, Fig. 3b), the diffusion of solute through the porous wall (red arrows) is fast compared to convection out of the phloem (blue arrows) i.e. the total hydraulic resistance is large. This leads to similar concentrations in the phloem and mesophyll (see concentration diagrams of Fig. 3b) and corresponds to Münch pumping:

$$c \approx c_0, \quad Q \approx RTc_0/R_{\text{tot}} \text{ and } \phi \approx RTc_0^2/R_{\text{tot}} \text{ for } f \ll 1. \quad (6)$$

In this regime, export is limited by the convection of water through the hydraulic circuit. Since the phloem concentration is close to that of the source, a plant working in this regime of low flushing number would benefit from the maximum pressure that the accumulated solutes can deliver and would be able to make effective use of its loading potential.

For large flushing numbers ($f \gg 1$, Fig. 3d), where the diffusive loading conductivity is small compared to the mean flow rate (i.e. the total hydraulic resistance is small) the diffusion of solute through the physical membrane (red arrows) is slow compared to convection out of the phloem (blue arrows). The concentration $c$ in the phloem is thus much smaller than the concentration in the source (see concentration diagrams in Fig. 3d). Expanding Eq. (4) for large $f$, we find:

$$c = c_0/\sqrt{f} = \sqrt{R_{\text{tot}} K_D c_0/RT} \text{ and } Q = Q_M/\sqrt{f} = \sqrt{K_D RT c_0/R_{\text{tot}}} \text{ for } f \gg 1. \quad (7)$$

In this regime, export is limited by the diffusion of solutes through the porous membrane and the water flow rate shows a weak (-½ power) scaling with the system hydraulic resistance (Fig. 3d). In addition, the total export of sugars is given by:

$$\phi = \phi_M/f = K_D c_0 \text{ for } f \gg 1. \quad (8)$$

which corresponds to maximal diffusive transport across the porous interface. Plants working in this regime can not make full use of their loading potential, as the large downhill concentration gradient between mesophyll and phloem reduces the available driving force for the flow, leading to small export compared to the Münch case.

The nonlinear scalings of flow rate $Q \sim R_{\text{tot}}^{-1/2}$ and sugar concentration $c \sim R_{\text{tot}}^{1/2}$ with the hydraulic resistance $R_{\text{tot}}$ in the diffusion-limited regime ($f \gg 1$) are unusual and reflect the dependence of the osmotic driving force on the flow that it creates. In the diffusion-limited regime, an increase of the system's hydraulic resistance leads to a build-up of the concentration $c$ in the phloem, which in turn increases the driving force for water flow (Fig. 3c-b). This diffusion-limited regime could present potential engineering applications in situation where steady flow or constant solute export is necessary, regardless of the output resistance or backpressure.

Finally, we note that the dimensions of the artificial phloem channels (~60 μm) are comparable in size to those of the largest sieve elements (~40 μm). In addition, flushing numbers achieved by the device – which are characterized by loading conductance, and hydraulic resistance – span botanically relevant regimes (see Fig. 5). Hence we expect similar effects to occur in real plants and in our tree-on-a-chip.

*Phloem Mechanical Pressure*

Passive phloem loading has, until recently, been thought insufficient to drive transport over long distances, because the passive coupling of the mesophyll and the phloem does not allow for an increase in the available driving pressure $\Delta p \sim RTc$ (5). More fundamentally, direct experimental evidence of the ability of the Münch mechanism to osmotically generate the very large hydrostatic pressures (on the order of several bars) necessary to drive long-distance transport in the phloem are scarce (29). To test these conjectures, we connected the device outlet to capillary tubes of various resistances (Figs. 1c-d) and measured the resulting flow rate $Q$ (Fig. 4a) and hydrostatic pressures $\Delta P$ (Fig. 4b) developed in the phloem. Measurements were made with a source concentration $c_0$ = 110 mmol/L corresponding to an osmotic pressure of approximately 12.7 bars which is similar to the mesophyll osmotic pressure in trees.

We estimate the steady state transport resistance as $R_T = \Delta P/Q$. This transport resistance can reach values as high as $10^{18}$ Pa.s/m³, equivalent to a continuous sieve tube of radius 20 µm and up to 10 meters in length (Fig. 4a, upper legend). As the transport resistance is increased, the flushing number decreases (Fig. 4b, upper legend) and we observe a decrease in the pumping speed (Fig. 4a) and an increase in the hydrostatic pressure (Fig. 4b) and phloem concentration (Fig. S3), in agreement with our analytical model (see Eq. (S11)).

Importantly, we find that phloem hydrostatic pressures in our synthetic passive phloem loader can reach values up to 10 bars. This value is close to the total osmotic pressure of the source, of the same order as the pressures needed to drive translocation in large trees (5, 30–32), and comparable to the pressures recently measured in the phloem of active loaders (29). Our synthetic tree-on-a-chip thus shows that passive loading mechanisms are able to generate the large osmotic pressures necessary to drive long-distance transport in plants.

*Meta-Analysis of phloem loading strategies in trees and herbaceous plants*

We now turn our focus to the export of sugar within the passive pumping processes elucidated in Figs. 3-4. The black curve in Fig. 5 presents the Münch efficiency (ratio of the export rate of sugar over the export rate in the Münch case, Eq. (6)) as a function of the flushing number $f$. Münch efficiency plateaus to a maximum of 1 at low flushing number (for which phloem and mesophyll concentration are equal), and decreases for larger flushing number, as phloem concentration decreases, following the same trend as normalized water export (Fig. 3a). We now seek to determine which transport regimes plants occupy, and how this regime depends on both global plants traits such as height, sieve tube radius and mesophyll osmotic potential and local traits such as plasmodesmata geometry and density. In the following paragraphs, we will estimate biologically relevant valued for each of these parameters.

If we assume that the pressure differential between leaf xylem and root scales with tree height as $P_X - P_R \approx \rho g h$, we can, for cylindrical sieve elements, approximate the flushing number as (from Eq. (3)):

$$f \simeq \frac{RTc_0 - \rho g h}{K_D(R_M + R_T)} = \frac{1}{k_D r_M} \frac{RTc_0 - \rho g h}{1 + \frac{16\eta h l}{r_M a^3}}. \qquad (9)$$

where $a$ is the phloem sieve tube radius, $l$ is the length of the loading zone (of the order of leaf length), $h$ is the length of the transport zone (of the order of tree height), and $\eta$ is viscosity of the phloem sap. We also introduce the hydraulic resistivity of the semipermeable membrane $r_M$ [Pa.s/m] and diffusive loading conductivity $k_D$ [m/s], which are related to the membrane resistance $R_M = r_M/A$ and diffusive loading conductance $K_D = Ak_D$ by the mesophyll-to-phloem contact area $A = 2\pi al$. We also assume that the transport resistance follows poiseuille law for viscous flow, and depends on plant height $h$ and sieve tube radius $a$ such that $R_T = 8\eta h/\pi a^4$.

Passive loaders can maximize their Münch efficiency at low flushing number by increasing the loading conductivity of their plasmodesmatal interface ($k_D$, Eq. 9). In the limit of very large conductivity $k_D$, phloem and mesophyll cells have similar concentrations. Thus, sugar export reaches the Münch sugar export rate $\phi_M$, the maximal export achievable for a given source concentration and transport resistance (Eq. 6). However, several factors might limit the conductivity of the mesophyll-phloem plasmodesmata, and could prevent all passive loaders from working in the very low $f$ regime. Here, we use a simple model to express plasmodesmatal interface conductivity in term of geometrical parameters as $k_D = \rho_P N \pi r^2 D/e$, where $\rho_P$ [m$^{-2}$] is the plasmodesmatal density, $N \approx 9$ is the number of channels in each plasmodesma, $r$ [m] is the radius of the individual plasmodesmatal channels, $D$ [m$^2$/s] is the effective sugar diffusivity and $e$ [m] is the length of the plasmodesmatal channels. Plasmodesmatal pores must allow small sugar molecules to diffuse, while maintaining the integrity of the mesophyll cells, by preventing larger structural proteins or organelles from passively leaving the cell. We thus expect the plasmodesmatal pore radii at the mesophyll/phloem interface to be larger than the sucrose hydrodynamic radius and smaller than the typical globular protein. Hence, we estimate plasmodesmatal pores to have radii around $r \approx 1$ nm, almost twice the size of sucrose. Plasmodesmatal channel lengths are approximated as equal to the cell wall thickness $e \approx 250$ nm, and we take a diffusivity $D \approx 4.7 \cdot 10^{-11}$ m$^2$/s for sucrose confined in plasmodesmatas *(see Materials and Methods)*. For passive symplastic loaders, the density of plasmodesmata at the mesophyll-phloem interface is limited to a maximal value of $\rho_P = 10$ to 40 plasmodesmata/µm$^2$ of interface (type 1 and 1-2a of Gamalei's classification (33)). Using these estimates, we can assume that the conductivity $k_D$ of the plasmodesmatal interface does not vary dramatically among passive symplastic loaders, and has an upper value of $k_D = 0.2$ µm/s (*Materials and Methods*). This value is of the same order of magnitude as experimentally determined loading conductivities (48).

Taking the plasmodesmatal conductivity $k_D$ as determined above, we plot in the inset of Fig. 5 the absolute export rate per surface area of the phloem, as a function of the flushing number. For fixed $k_D$, export increases with increasing convection in the phloem, i.e. larger flushing number, despite a decrease in Münch efficiency.

Among the other physiological parameters of Eq. (9), which might influence flushing number in real plants, the semipermeable membrane resistance $r_M$ between xylem and phoem appears to be slightly larger in trees than small plants (34, 35). This trend will amplify the fact that larger plants have larger total transport resistances, which is already taken into account in (Eq. 9) via the dependence of flushing number on height. For simplicity, we thus take in the following a fixed value $r_M \approx 2 \cdot 10^{13}$ m/s/Pa for the membrane hydraulic resistance. Sap viscosity $\eta$ has been reported to be relatively constant across species (36), and we take here an effective viscosity $\eta \approx 5$ mPa.s, accounting for additional sieve plates resistance (37). We also take fixed mesophyll osmotic potential $RTc_0 \approx 10$ bars, assuming relatively constant mesophyll osmotic pressure among passive loaders.

The remaining parameters, loading phloem length $l$, transport phloem length $h$, and phloem sieve element radius $a$, relate to macroscopic traits, more readily measurable in a broad set of plants. Loading phloem length will be of the order of leaf length $l$, which varies considerably by species and by height (38). For trees taller than 20 m, the sieve element radius $a$ does not seem to grow larger than 20 μm (38). Transport phloem length, of the order of plant height $h$ will thus play a critical role in setting the flushing number, by both increasing transport resistance ($R_T$ in the denominator of Eq. (9)) and decreasing the available water potential difference between xylem and mesophyll ($RTc_0 - \rho g h$, in the numerator). Although simplified, our assumptions provide a first basis for exploring trends in the flushing number with plant traits.

Based on these considerations, we evaluate the flushing number in a set of real plants by taking the macroscopic values for height $h$, leaf length $l$ and sieve tube radius $a$ from Jensen et al. (39). The box and whiskers plots in Fig. 5 represent the distribution of flushing numbers for plants that are thought to load passively (angiosperm trees in blue and gymnosperm trees in green) and actively (herbs in red). The extent of each distribution is reported in the colored domains. Central lines and dots indicate distribution median. Due to their large height, the distribution of both tree datasets (blue and green) is skewed towards lower flushing number compared to herbaceous plants (red) and a large fraction of passive loaders exist on the export plateau at lower $f$. These large plants suffer from reduced export rate (Fig. 5, inset), but this reduction is intrinsic to the physiological constraints set by their large transport resistances and low xylem pressure, and should affect all species regardless of their loading mechanisms. On the other hand, those large passive loaders are seen to operate at maximal Münch efficiency, where export is very close to the Münch export rate and where phloem concentration and source concentration are approximately equal, in agreement with experimental observations (7, 40).

**Discussion**

Fu et al. (9) suggested that large plants must have a large sugar concentration in their leaves so as to offset low leaf xylem pressure. Using the meta-analysis summarized in Fig. 5, we build upon this hypothesis to show that tall passive loaders (angiosperm and gymnosperm tree, blue and green datasets in Fig. 5) can take full advantage of their high mesophyll concentration. Due to their large height, they can sustain small concentration gradients between mesophyll and phloem, and function as perfect Münch-like pumps. Our experiments also demonstrate that passive loaders can develop sufficient pressure to drive long-distance phloem transport, with no need to invest metabolic energy in active mechanisms (Fig. 4b). We thus argue that passive loading represents an effective export strategy for large trees, which are required to maintain large sugar concentration in their mesophyll cells.

Although sieve element radius is generally smaller for plants of smaller height, we observe in Fig. 5 that if herbs were to depend on a passive loading process, they would operate at larger flushing number, principaly owing to their lower transport resistance and larger xylem water potential (red distribution). This situation would lead to diffusion-limited export, lower Münch efficiency and storage of large sugar concentrations in mesophyll cells compared to the phloem, a situation detrimental for growth in herbs (1). This observation suggests that active modes of sugar transfer between mesophyll and phloem may have emerged in herbs to overcome this diffusion limitation and reduce the amount of sugars stored in the mesophyll cells. Similarly, some of the tree species operating at larger flushing number may have developed active stragegies to complement or replace passive loading strategies and overcome diffusion-limited transport through their plasmodesmata.

It is important to note that we have made a number of idealizations in our analysis. First we considered the idealized situation where the dominant sugar gradient lies at the mesophyll-phloem interface. Because transport between mesophyll cells is also symplastic and passive, additional concentration gradients along mesophylls cells and in the pre-phloem pathway could lead to a potential increase of the flushing effects (41). Second, we neglected in our analysis the possible coupling of the xylem with the mesophyll source cells, so as to simplify the transport equations and the experimental realization of the synthetic system. The existence of such direct coupling in plants is unclear, and it has been suggested that most of the hydrostatic pressure could be generated in the source itself, leading to both convective and diffusive transport through the plasmodesmata (41). In this situation, the flushing effects described herein still occur in the case of hindered plasmodesmatal transport (see *SI 3*).

As a next step, the relevance of the flushing number could be assessed experimentally in passive loaders. We predict than an increase in transport resistance, for example via the application of cold to inhibit long-distance transport, should lead to a direct build-up of phloem and mesophyll concentration and reduce the magnitude of the concentration gradient. Fu et al. (8) showed that the few herbs that load passively have reduced whole plant water conductivity. It would be interesting to see whether passive loading in herbs also correlates with the specific macroscopic traits predicted by our analysis such as smaller sieve tube radii and enhanced plasmodesmatal permeabilities. Arabidopsis with downregulated sucrose transporters was found to be able to complete its life cycle (42). One could check whether phloem loading could occur symplastically in Arabidopsis, for which plasmodesmata are present at all interfaces between the mesophyll and phloem and see whether the downhill concentration gradient between mesophyll and phloem is larger than in putative passive loaders.

Several explanations have been proposed to rationalize the use of different phloem loading strategies according to plant traits. In particular, Fu et al. (8) proposed that large sugar concentrations in the leaves of tall trees are required to sustain low xylem tensions. We showed here that the regime of low flushing numbers, for which Münch efficiency is maximal, is accessible to plants which load symplastically by diffusion provided the conductivity of the plasmodesmata is large enough compared to the mean flow rate in the plant. Based on physiological data, we showed that most trees are expected to work in such regimes, where the concentration gradient between source cells and the phloem is shallow, as confirmed by experimental measurements (7, 40). Because the hydrostatic pressures - of up to 10 bars - which can be developed in our synthetic system are comparable to the pressure required to drive translocation in large trees, we argue that active phloem loading is not required for long distance transport in plants. Tall passive loaders like trees can thus take full advantages of their high mesophyll concentration for export and develop hydrostatic pressures comparable to those observed in active loaders. Our analysis thus provides a new and unexpected correlation between macroscopic plant traits and phloem loading strategies.

## Materials and Methods

### Experiments
*Microfluidic osmotic pump*
The microfluidic osmotic pump was made by sandwiching together gaskets, channels and membranes (Figs. 1c-d). The part of the phloem channel interacting with the sugar and water source was 1 mm wide by 5.2 cm long. Gaskets were craft-cut out of 100 μm thick soft Polyvinylchloride (PVC) films. The phloem channel was craft-cut out of 60 μm thick rigid Polyester (PET) sheets, and its thickness increased up to 600 μm using additional plastic sheets. The water and sugar source channels were laser-cut out of 1 mm thick acrylic sheets, sand-papered and polished (Novus Plastic). The device was screwed together between aluminum plates. Flexible capillary tubes were connected to the pump channels using nanoports (Idex H&S; www.idex-hs.com) glued to the acrylic channel with cyanoacrylate glue (Super Glue Corp.) and sodium bicarbonate as an accelerator.

*Semipermeable membrane, porous wall and sugars*
The solute used was 6 kDa Dextran (Alfa Aesar). The solutions where prepared in 0.05% sodium azide aqueous solutions to discourage bacterial growth. The membranes used were cellulose ester (CE) dialysis membranes of various molecular weight cut-off (MWCO) (SpectrumLabs) and thickness approximately 60 μm. We used semipermeable membranes of respective MWCO 1 kDa (Fig. 2, green points) and 3-5 kDa (Fig. 2, red points, larger MWCO values lead to increasing water permeabilities) and physical membranes of MWCO 8-10 kDa. To attain the regime of large flushing numbers, $K_D$ was increased by layering several physical membranes together (Fig. 2, red triangles) or increasing the thickness of the phloem channel (Fig. 2, green triangles).
Resistance of the semipermeable membranes was measured independently by flowing dextran solutions of known concentration on one side of the membrane, DI water on the other side, and measuring the resulting initial flow rate. Using the less permeable membrane, for which solute depletion close to the membrane is negligible, we find a linear variation of flow rate with concentration. The resistance of this membrane was measured independently using sucrose solutions of known osmotic coefficient $\alpha = 1$, from which we could estimate the osmotic coefficient of dextran as $\alpha = 4.7$.

*Flow rate measurement*
Flow rates were measured by connecting the inlet and outlet of the pump to two partially filled glass capillaries of diameters 500 μm and 800 μm. For a quasi-static meniscus, the capillary pressure due to the water meniscus in the glass capillary is the same in the inlet and outlet of the pump. Monitoring the evolution of the meniscus position over time by time-lapse photography allowed us to calculate the flow rates $Q_{\text{inlet}}$ and $Q_{\text{outlet}}$. A measurement was considered valid when the absolute difference between inlet and outlet flow rates differed by less than 0.5 nL/s. This difference was attributed to the evaporative flux of water $Q_{\text{evap.}}$ at the two menisci interfaces. The pump flow rate was thus estimated as $Q_{\text{pump}} = (Q_{\text{inlet}} + Q_{\text{outlet}})/2$.
Due to the large size of the channel reservoir (around 16 cm$^3$, 50 times larger than the phloem channel), the source concentration did not vary much over the time of the experiments, and steady flow rate could be observed for several hours (Fig. 2, inset). When the MWCO of the semipermeable membrane was close to the dextran size (semipermeable membrane of 3.5-5 kDa

MWCO) and the concentration in the phloem channel was large, we observed a decrease in water flow approximately one hour after the establishment of a steady flow. This decrease was attributed to the diffusive leakage of solute to the xylem channel through the semipermeable membrane.

*Pressure measurement*
To allow the development of high pressures in the pump, the pump outlet was connected to capillary tubes (Polymicro Technology) of radius 15 and 10 μm and lengths ranging from 5 to 40 cm. At the beginning of a pressure measurement experiments, we flush the capillary tube, phloem and xylem with pure water and fill the source with a fixed concentration $c_0 = 110$ mmol/L. Due to the build-up of pressure, the phloem channel dilates and we observe initially a larger flow rate in the inlet than in the outlet. We then let the pump run for several hours to reach a steady state where phloem and capillary tubes are filled with the same concentration $c$, and the transport resistance $R_T$ remains constant. Once this steady state has been reached, we flush the xylem channel with pure water, refill the sugar reservoir, and carry out the experiment again. This time, pressure builds up more rapidly, as the capillary tube is already filled with the viscous dextran solution. Pressures are directly measured using two pressure sensors of respective range 0 to 6.9 bar and 0 to 17.2 bar (Honeywell 26PCFFM6G and 26PCGFM6G), which can be directly screwed to one end of the phloem channel (Fig. 1d, white dashed box).

To estimate the concentration of the dextran solution flowing in the capillary tubes (Fig. S3), we estimate the viscosity of the solution flowing in the tube by comparing the value of the estimated hydraulic resistance $R_T = \Delta P/Q$ with the expected poiseuille resistance calculated from the length and radius of the capillary tubes. The variation of the sugar viscosity with concentration was measured using a Cone and Plate rheometer, from which we could back-calculate solute concentration in the capillaries.

**Passive loading and plant data**
*Parameter estimation*
The diffusive loading conductance $K_D$, that characterizes the solute permeability of plasmodesmata at the mesophyll-phloem can be written as $K_D = Ak_D$ with A the contact area and $k_D = N\rho_P \pi r^2 D/d$ with d ≈ 0.25 μm the typical length of a plasmodesmata, $\rho_P \approx 40/\mu m^2$ the density of plasmodesmata, $D$ the effective diffusion coefficient of sucrose through the pores, N = 9 the number of pores per plasmodesmata and r ≈ 1 nm the radius of one pore. Assuming hindered sucrose transport (43), we obtain $D \approx 4.7 \cdot 10^{-11}$ m$^2$/s, leading to $k_D = 0.2$ μm/s.

To estimate the dependence of the flushing number on macroscopic plant traits, we approximate the mesophyll-phloem contact area as $A = 2\pi a l$ with *a* the mean sieve element radius, *l* the leaf length and *h* being plant height. We thus have $R_M = 1/(2\pi a l \, L_P)$ and $R_P = 8\eta h/(\pi a^4)$, leading to Eq. (9). The flushing number can then be estimated from (39) using only the macroscopic traits *a*, *l* and *h*.

**Data availability**
The data that support the findings of this study are available from the corresponding author upon request.

**Correspondence**

Correspondence and request for materials should be addressed to Jean Comtet: jean.comtet@gmail.com.

**Acknowledgement**

Jean Comtet would like to thank José Alvarado for his support, Alexander Barbati for advice on the device fabrication and Bavand Keshavarz for help with the rheology measurements.

A.E.H. and J.C. acknowledge support from DARPA (W31P4Q-13-1-0013). K.H.J. was supported by a research grant from VILLUM FONDEN (13166). A.D.S. acknowledges support from the AFOSR (FA9550-15-1-0052). R.T. was supported by National Science Foundation (USA) (grant no. IOS-1354718).

J.C. and A.E.H. conceived the project. J.C. designed and executed the experiments and developed the theoretical model, with input from all authors. J.C., K.H.J., R.T., A.D.S. and A.E.H. interpreted the experimental data and the meta-analysis and wrote the paper.

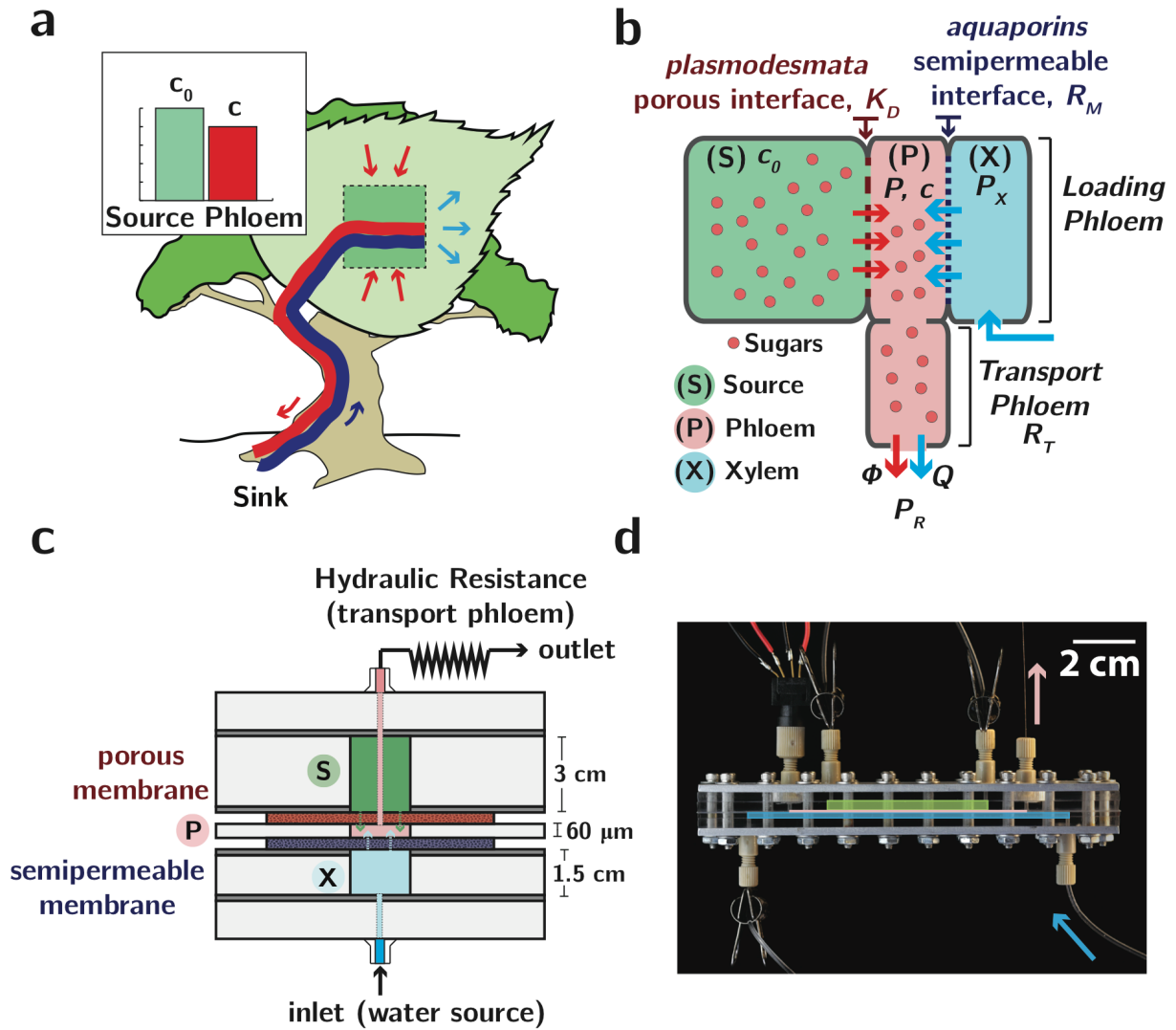

**Fig. 1. Passive phloem loading in plants and in the synthetic tree-on-a-chip. (a)** Schematic of a tree, where xylem (blue line) pulls water to the leaves via evaporation (blue arrows), and phloem (red line) exports the photosynthesized sugars (red arrows) from sources (green) to sinks. **(b-c)** Simplified model for water and sugar transport in passive symplastic loaders and in our synthetic tree-on-a-chip. Sugars (red dots), stored in the source (S, green) diffuse to the phloem (P, pink) through a porous wall, where they drive, by osmosis through a semipermeable membrane a flow of water from the xylem (X, blue). Water and sugars are subsequently convected down to the roots through the transport phloem. **(c)** Cross-sectional view of the osmotic pump and **(d)** photograph of the device.

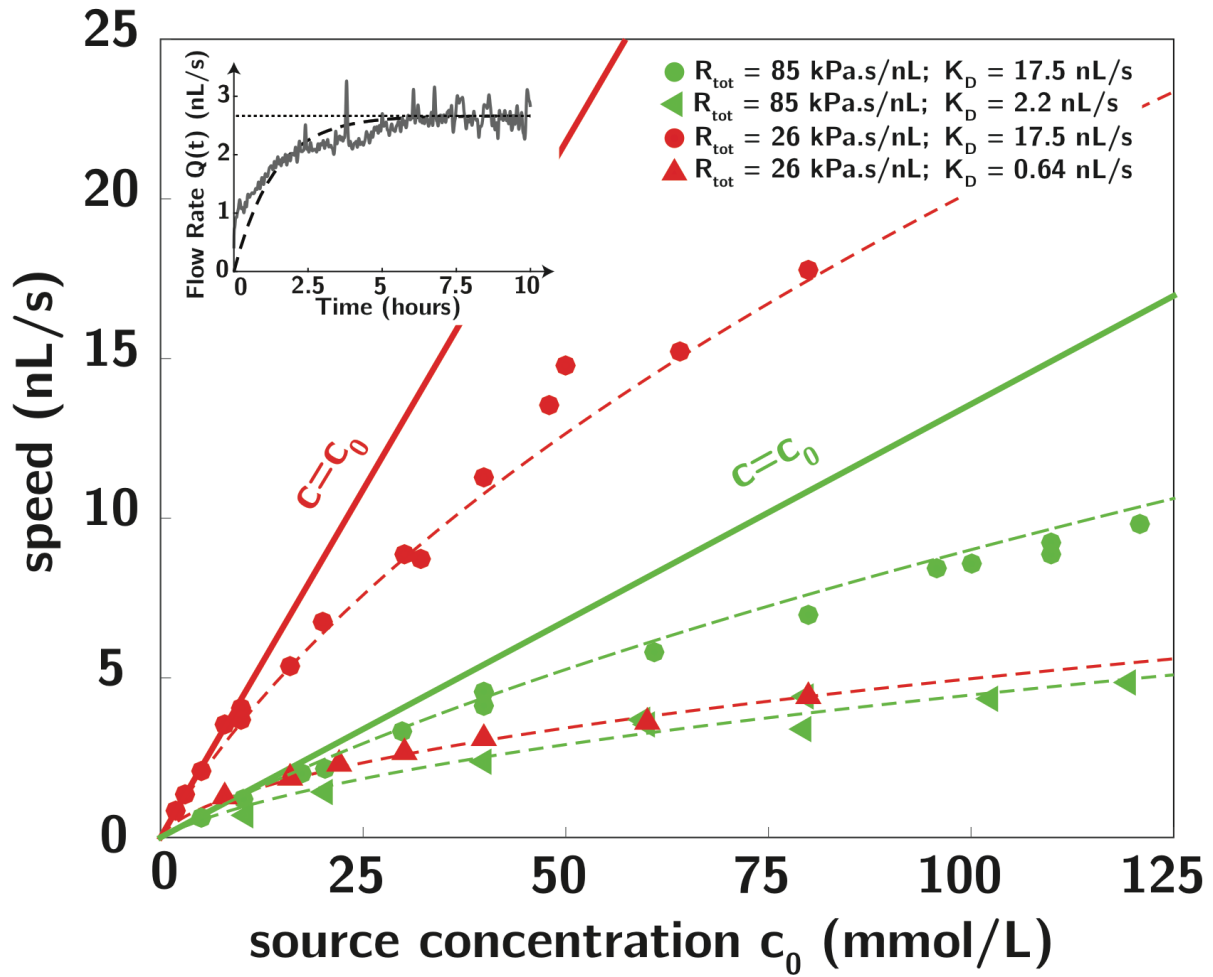

**Fig. 2. Steady-state flow rates deviate from standard Münch models.** Steady-state flow rate as a function of source concentration $c_0$, for different values of the total hydraulic resistances $R_{tot}$. Dotted lines are fits to Eq. (4), allowing the determination of the porous wall loading conductivity $K_D$. Straight plain lines represent the expected flow rate $Q_M = RTc_0/R_{tot}$ from the Münch model where phloem concentration $c$ is equal to source concentration $c_0$. Inset: for a given source concentration, flow rate $Q(t)$ reaches a steady-state value where sugar loading via diffusion and sugar export via convection balance (horizontal dashed line).

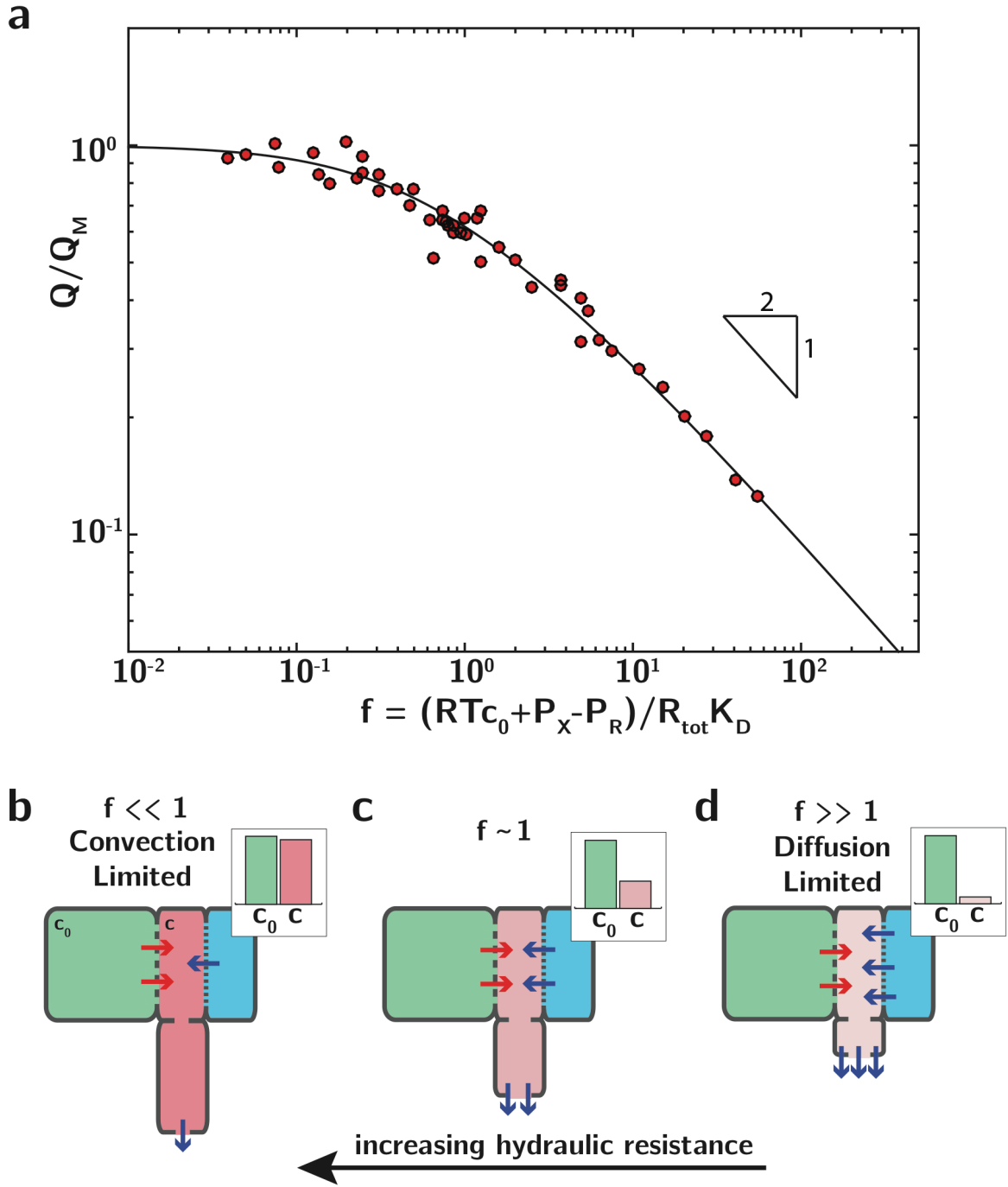

**Fig. 3. Convection and diffusion limited export regimes. (a)** Variation of the dimensionless flow rate $Q/Q_M$ with the flushing number. The experimental points correspond to the data points of Fig. 2. The solid line shows the solution of Eq. (4). **(b-d)** Schematic representation of the regimes of low (Eq. (6)) **(b)**, intermediate **(c)** and large (Eq. (7)) **(d)** flushing number. Increase in the flushing number leads to increase in convective export (blue arrows) compared to diffusive sugar loading (red arrows), and decrease in phloem concentration compared to that of the source (concentration diagrams).

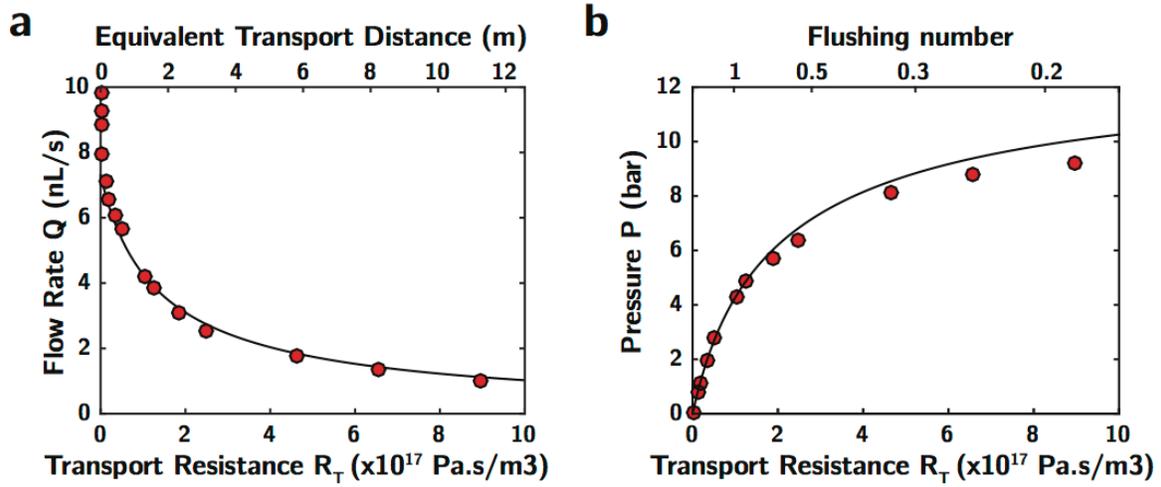

**Fig. 4. Large hydrostatic pressures can be obtained in the synthetic passive loader.** Variation of flow rate **(a)** and phloem hydrostatic pressure **(b)** with transport resistance (or equivalent transport distance, upper legend) for a fixed source concentration corresponding to an osmotic pressure of 12.7 bar. Plain lines correspond to the theoretical expression from Eqs. (4) and (S11) with a mass tranfer coefficient $K_D$ = 7 nL/s, and membrane resistance $R_M$ = 85 kPa.s/nL.

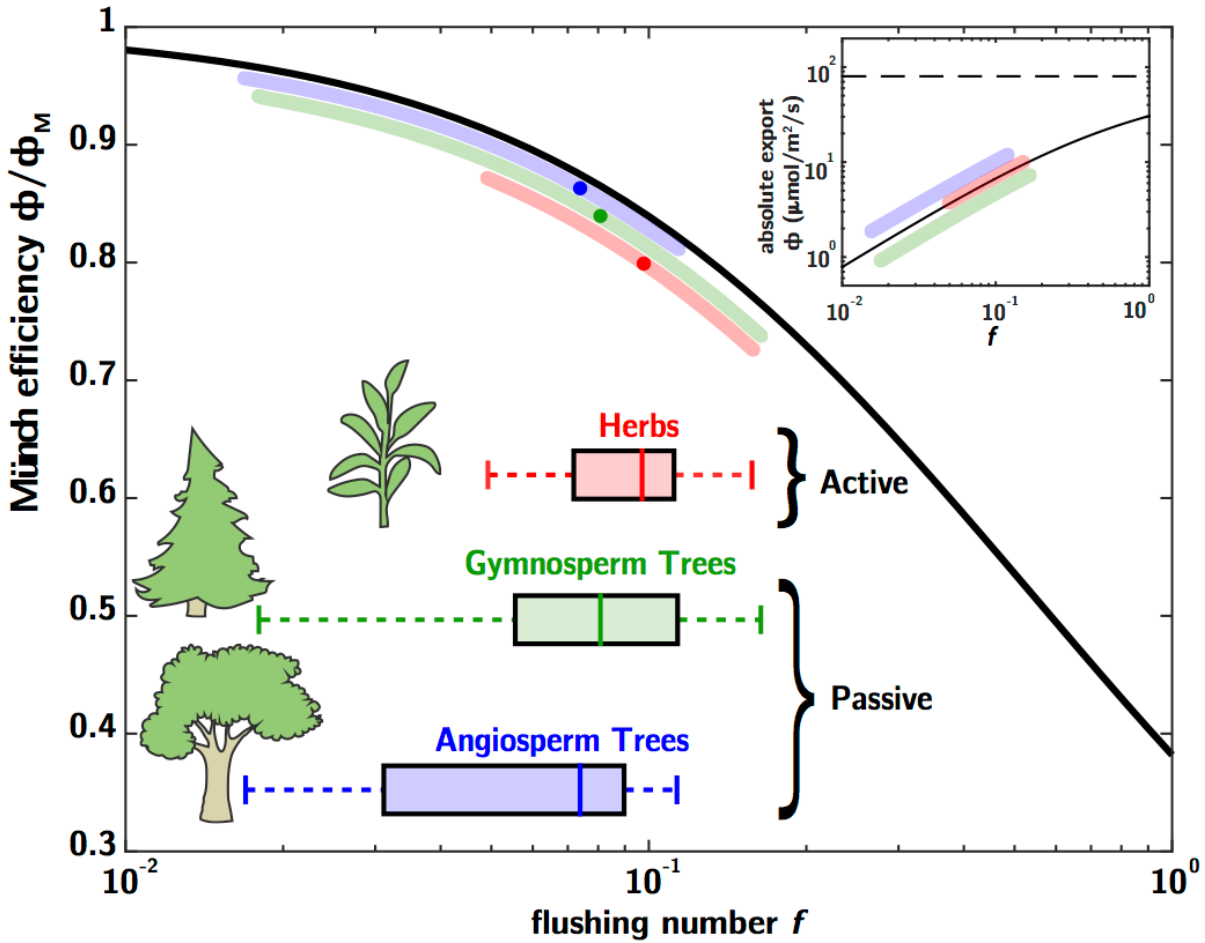

**Fig. 5. Meta-analysis of phloem loading strategies in trees and herbs.** Münch efficiency (Eq. 5) decreases with increasing flushing number $f$. Box-and-Whisker plots: estimation of the flushing number for a sample of gymnosperm and angiosperm trees and herbs. Central line indicates distribution median, 50% of the data points are inside the box and whiskers indicate extremums. Colored domains reproduce flushing number distribution, with dots indicating distribution median. Inset: absolute export rate per unit area of the phloem for a fixed plasmodesmatal permeability $k_D = 0.2$ µm/s, in terms of the flushing number.

| Symbol | Definition [unit] |
|---|---|
| $f = \dfrac{RTc_0 + P_X - P_R}{R_{tot} K_D}$ | Flushing number, the ratio of advective export to diffusive loading in passive phloem loading |
| $c_0$ | Source (Mesophyll) concentration [mmol/L] |
| $c$ | Phloem concentration [mmol/L] |
| $R_T$ | Transport phloem resistance [Pa.s/m$^3$] |
| $R_M$ | Loading phloem resistance [Pa.s/m$^3$] |
| $R_{tot} = R_M + R_T$ | Total hydraulic resistance [Pa.s/m$^3$] |
| $K_D$ | Diffusive loading conductance [m$^3$/s] |
| $Q$ | Water flow rate [m$^3$/s] |
| $Q_M = \dfrac{RTc_0 + P_X - P_R}{R_{tot} K_D}$ | Münch water flow rate [m$^3$/s], obtained for equal phloem and mesophyll/source concentration |
| $\phi$ | Sugar export rate [mol/s] |
| $\phi_M = Q_M \cdot c_0$ | Münch sugar export rate [mol/s], obtained for equal phloem and mesophyll/source concentration |
| $P_X$ | Xylem Pressure [Pa] |
| $P_R$ | Root Pressure [Pa] |
| $P$ | Phloem Pressure [Pa] |

**Table 1:** Description of symbols used.